# Dynamics of the Ride-Sourcing Market: A Coevolutionary Model of Competition between Two-Sided Mobility Platforms

Farnoud Ghasemi*[1], Arkadiusz Drabicki[2], Rafał Kucharski[3]

[1] PhD Candidate, Mathematics and Computer Science, Jagiellonian University, Poland

[2] PhD Candidate, Department of Transport Systems, Cracow University of Technology, Poland

[3] Assistant Professor, Jagiellonian University, Poland

**SHORT SUMMARY**

There is a fierce competition between two-sided mobility platforms (e.g., Uber and Lyft) fueled by massive subsidies, yet the underlying dynamics and interactions between the competing platforms are largely unknown. These platforms rely on the cross-side network effects to grow, they need to attract agents from both sides to kick-off: travellers are needed for drivers and drivers are needed for travellers. We use our coevolutionary model featured by the S-shaped learning curves to simulate the day-to-day dynamics of the ride-sourcing market at the microscopic level. We run three scenarios to illustrate the possible equilibria in the market. Our results underline how the correlation inside the ride-sourcing nest of the agents choice set significantly affects the platforms' market shares. While late entry to the market decreases the chance of platform success and possibly results in "winner-takes-all", heavy subsidies can keep the new platform in competition giving rise to "market sharing" regime.

## 1. INTRODUCTION

Ride-sourcing companies such as Uber and Lyft have achieved significant market share in a short time through the two-sided platform business model. The reason underlying such a tremendous potential to grow in two-sided markets is the power of network. The classic definition by Rochet and Tirole (2006), characterizes them as the markets in which one or several platforms enable interactions between end-users and try to get both sides on board by appropriately charging each side. The platforms associated with these markets rely on the critical mass required for their self-sustainable operations and the network effects to induce growth (Belleflamme & Peitz, 2016). Platforms apply various market entry strategies in the early adaptation phase to follow a desired growth pattern that includes different stages from their launch up to maturity.

Even though ride-sourcing platforms have the potential to grow rapidly, they fiercely compete over the common pool of travelers and drivers. Considering the so-called multi-homing characteristics of the market, in which users can move from one platform to another with ease, it becomes extremely challenging for the platform to gain and hold the market share. As for the platforms new to the market, they need to first induce the interactions between the decentralized supply and demand to reach market shares sufficient to trigger the cross-side network effects. On the other side, existing platforms with stable market share adjust their strategies in accordance with other platforms to avoid losing market share. These platform strategies are mostly controlled by subsidies and implemented through e.g., discounts, and incentives.

There is a wide body of research on the competition between the platforms. Relying on the game theory, Zhang and Nie (2021) study ride-sourcing market in which two platforms compete with each other, as well as with transit, and Ahmadinejad et all., (2019) examine the competition impact on ride-sourcing parties by adjusting the trip fare. Using analogous methodology, Siddiq and Taylor (2022) throw light on the importance of autonomous vehicles for platforms' profitability. In different approaches, Shoman and Moreno (2021), conduct a stated preference analysis to find the ride-sourcing impact on the modal split in city of Munich.

*Study approach and contribution*

Previous studies are either equilibrium-based or assume fixed demand and/or supply and they neglect the interactions between the parties driving the complex system evolution. Here, we illustrate with our experiments, an adequate framework to realistically model the platform competition in two-sided model with subsidizing strategies is missing.

Previously, we proposed a novel, microscopic co-evolutionary model which is capable of reproducing platform's growth mechanism day-to-day. The key element of the model is the S-shaped learning curves which enable the agents to adapt and stabilize their behavior and yet to remain sensitive to the changes in their environment (Ghasemi and Kucharski, 2022). In this study, we extend the previous model with platform competition considering the multi-homing characteristics of the system. We incorporate nested choice modelling to examine the correlation between platforms and the possible equilibria in the ride-sourcing market.

## 2. METHODOLOGY

We model two-sided mobility market with MaaSSim[1] agent-based simulator (Kucharski and Cats, 2022), extended here with a coevolutionary model to represent the day-to-day dynamics of two-sided mobility market. We simulate two classes of agents representing two sides of the system and a platform as an intermediate agent matching the demand to the supply. A pool of travelers and drivers, who are not formerly notified about our ride-sourcing platform, gradually become aware of the ride-sourcing. When an agent gets notified, he/she may decide to participate in the market – i.e., supply the demand as a driver or travel to his/her destination as the platform client. With the participation, agents start to learn and adapt their behavior through endogenous and exogenous factors.

*Platform*

Platform executes the strategy $S_t$ with the control levers, namely: trip fare $f_t$, commission rate $c_t$, discount $d_t$ and marketing $m_t$ for each day $t$ of the simulation.

$$S_t = \{f_t, c_t, d_t, m_t\} \tag{1}$$

*Traveller*

Each notified traveler $r$ on day $t$ selects between alternatives from the choice *set* $C_r = \{rs, pt\} = \{\{p_1, p_2\}, pt\}$ including public transport ($pt$) and two ride-sourcing ($rs$) platforms ($p_1, p_2$). While the utility of public transport is fixed (formulated with a typical access/egress, waiting times, transfers, etc.), the platforms' utility is composed of multiple components, each adjusted day-to-day (as detailed in the upcoming sections).

---
[1] https://github.com/RafalKucharskiPK/MaaSSim

*Driver*

Analogous to travellers, each notified driver $d$ makes a choice from the choice set $C_d = \{rs, pt\} = \{\{p_1, p_2\}, rw\}$ which includes: working for ride-sourcing ($rs$) platform one ($p_1$) and platform two ($p_2$), and working elsewhere, for a fixed reservation wage ($rw$). The utility of working for platform as a driver is composed of same components of travellers and adjusted day-to-day as detailed below.

*Choice Utility*

For any notified agent $i$, we propose the generic perceived utility ($U$) formulation composed of three components: experience ($U^E$), marketing ($U^M$) and word of mouth ($U^{WOM}$):

$$U_{i,t} = \beta_i^E \cdot U_{i,t-1}^E + \beta_i^M \cdot U_{i,t-1}^M + \beta_i^{WOM} \cdot U_{i,t-1}^{WOM} + ASC + \varepsilon_i \qquad (2)$$

Agents every day ($t$) choose based on experiences collected until the previous day ($t$-1). Experienced utility is endogenous and comes directly from the simulation: drivers experience the actual incomes and operating cost, travelers experience travel time, waiting time and trip fare. Marketing is an exogenous factor being positive or negative (e.g., recent Uber scandal). Word-of-mouth is shared among agents over the social network. The $\beta$'s in the formula reflect the relative weights of respective utility components (ensuring that $\beta_i^E$, $\beta_i^M$, $\beta_i^{WOM} > 0$ and $\beta_i^E + \beta_i^M + \beta_i^{WOM} = 1$). The $ASC$ captures the effect of unobserved factors on the perceived utility of alternatives and $\varepsilon_i$ is the random utility error term. In such form, the utility is consistent with the discrete choice theory and can be applied e.g., in the logit model.

*S-shaped learning and adaptation*

The key element of the proposed model is the following adjustment mechanism which allows us to realistically represent the agents' dynamics specific to the platform growth. Agents learn and adapt their choice day-to-day based on the perceived utility components. Here, instead of exponential memory curve (used e.g., in de Ruijter et al., 2021), we follow Murre (2014) and propose a more adequate formulation of the so-called S-shaped learning curve in the context of urban mobility (Ghasemi and Kucharski, 2022). Fig. 1 provides a basic idea of our model.

The adjustment process can be seen as moving each of the utility components along the S-shaped curve with each of the utility components. Positive experience increases the experienced utility pushing the perception towards upper tail, and negative experience decreases it pushing the perception towards lower tail. The two extreme points (lower and upper endings) of curve represent absolutely negative and positive attitudes and learning can go both directions on any day. Triggered by the consecutive positive/negative experiences, learning proceeds slowly for the agents who already have sharp, extreme opinions, and is fast for the neutral agents. To this end, on the contrary to state-of-the-art models, we can stabilize the agents' behavior and, at the same time, remain sensitive to the system changes.

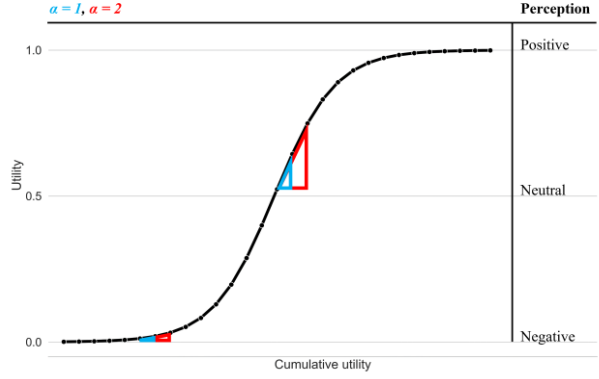

**Figure 1:** S-shaped curves used for the day-to-day learning process. The adjustment volume not only depends on sensitivity $\alpha$ but also the signal strength ($\Delta u$) and the position on the S-shaped curve.

Technically, we obtain the utility of respective components $c \in \{E, WOM, M\}$ after day $t$ ($U_{i,t}^c$) as follows. First, we retrieve the cumulative utility on the previous day $t-1$ by applying the inverse sigmoid function (eq. 3). Then, we update it with the difference coming from today $t$ (eq. 4), i.e. the signal strength. We weight it with the learning speed parameter $\alpha$ (determining step size on the S-shaped curve). Eventually, to obtain the updated utility at the end of day $t$ we use sigmoid (logistic) function with shape parameter $\beta$ (5):

$$CU_{i,t-1}^c = ln(\frac{1}{U_{i,t-1}^c} - 1) \tag{3}$$

$$CU_{i,t}^c = CU_{i,t-1}^c + \alpha . \Delta u_{i,t}^c \tag{4}$$

$$U_{i,t}^c = \frac{1}{1 + exp(\beta . CU_{i,t}^c)} \tag{5}$$

The above formulation is generic to represent various kinds of learning new experiences and exposure to effects. For the purpose of this study, we introduce specific formulas for three components of utility.

For the experience we adjust the utility as follows. Experienced cumulative utility of drivers $d$ on day $t$ for platform $p$ is updated with the relative difference between the reservation wage ($RW_d$) and the income experienced on that day (eq. 6). Similarly, traveler $r$ adjusts his/her experienced cumulative utility on day $t$ (eq. 7) according to relative difference between the experienced utility of the platform $p$ (as a function of waiting time, travel time, and trip fare) and the public transport ($pt$).

$$\Delta u_{d,t}^{E,p} = \frac{RW_d - E_{d,t}}{RW_d} \tag{6}$$

$$\Delta u_{r,t}^{E,p} = \frac{U_r^{pt} - E_{r,t}}{U_r^{pt}} \tag{7}$$

The marketing spreads uniformly among all the agents (target clients) and accumulates in time over the period of the marketing campaign. While marketing is constant before and after the campaign, it produces a positive effect on each exposure. The chance of agent $i$ to be exposed to the marketing on day t depends on the campaign intensity ($p_i^M$, from [0,1] range). We update the cumulative utility for marketing as follows:

$$\Delta u_{i,t}^{M,rs} = p_i^M (U_{i,t}^M - 1) \tag{8}$$

For the word-of-mouth, we assume pairwise interactions through the social network with agents, who share their perceived utility with each other. Analogously to the marketing, the WOM intensity ($p_{i,j}^{WOM}$) determines the likelihood of agent $i$ to share his/her opinion with agent $j$ on day $t$. Influenced by the exchange of views with their peers, agents adjust their cumulative utility of word-of-mouth as follows:

$$\Delta u_{i,t}^{WOM,rs} = p_{i,j}^{WOM} (U_{i,t}^{WOM} - U_{j,t}) \tag{9}$$

*Participation probability*

An agent starts considering a ride-sourcing platform in her mode choice set ($C_i$) only after being notified about it. Due to correlation between the ride-sourcing platforms, we apply the nested logit model. We assume the agent first selects between the alternative ($pt/rw$) and ride-sourcing, and when she selected ride-sourcing she choose among the competing platforms that she is notified about. The participation probability of notified agents is updated every day and depends on the perceived utility of alternatives as follows. The probability of choosing alternative $k$ (eq. 13) is the product of probability of $k$ inside the nest (eq. 10) and the probability of nest $n$ (eq. 12) based on the expected maximum utility of nest ($W^n$). $I_{i,t}^k$ is a binary variable switching from zero to one when agent gets notified about alternative k. The scale parameters are $\theta$ (at the upper choice level) and $\theta_n$ (within the ride-sourcing nest) which allows us to calculate the correlation inside the nest ($\rho \in [0,1]$) as: $\rho = 1 - \frac{\theta_n}{\theta}$.

$$P_{i,t}^{k/n} = I_{i,t}^k \frac{exp(\frac{U_{i,t}^k}{\theta_n})}{\sum_{k' \in K} exp(\frac{U_{i,t}^{k'}}{\theta_n})} \tag{10}$$

$$W_{i,t}^n = \theta_n . log(\sum_{k' \in n} exp(\frac{U_{i,t}^{k'}}{\theta_n})) \tag{11}$$

$$P_{i,t}^n = \frac{exp(\frac{W_{i,t}^n}{\theta})}{\sum_{n' \in N} exp(\frac{W_{i,t}^n}{\theta})} \tag{12}$$

$$P_{i,t}^k = P_{i,t}^{k/n} . P_{i,t}^n \tag{13}$$

*Experimental design*

We experiment on Amsterdam, with 2000 travelers and 200 drivers. The reservation wage of drivers is assumed 10.63[€/hour]. The operational costs (fuel, depreciation costs, etc.) of the drivers amount to 0.25 [€/km]. Each run simulates 4 hours of interactions between the parties. Vehicle speed is set to the flat 36 [km/h]. We consider the ride-sourcing fare of 1.2 [€/km] with a minimum of 2 [€] (based on the Uber price estimator). The utility weights of three main component are fixed as: $\beta^E = 0.7, \beta^{WOM} = 0.2$, and $\beta^M = 0.1$, while Marketing and WOM intensity are set to $p_i^M = p_{i,j}^{WOM} = 10\%$. We assumed the value of time 10.63 [€/hour] to compute the experienced utility of travellers.

## 3. RESULTS AND DISCUSSION

We first illustrate how the single platform competes against public transport in the period of one year (Figure 1). As agents get notified, they are initially reluctant to select the new travel mode. Yet, once travellers try out the platform and experience its benefits (40% discounts), they start to adapt and use ride-sourcing more frequently. This provides adequate income for the drivers. As the network gets denser on both demand and supply sides, it generates greater cross-side network effects. These effects provide both travellers and drivers with extra utilities in terms of lower waiting times and higher incomes, respectively. Thanks to this, the system grows and stabilizes around day 200.

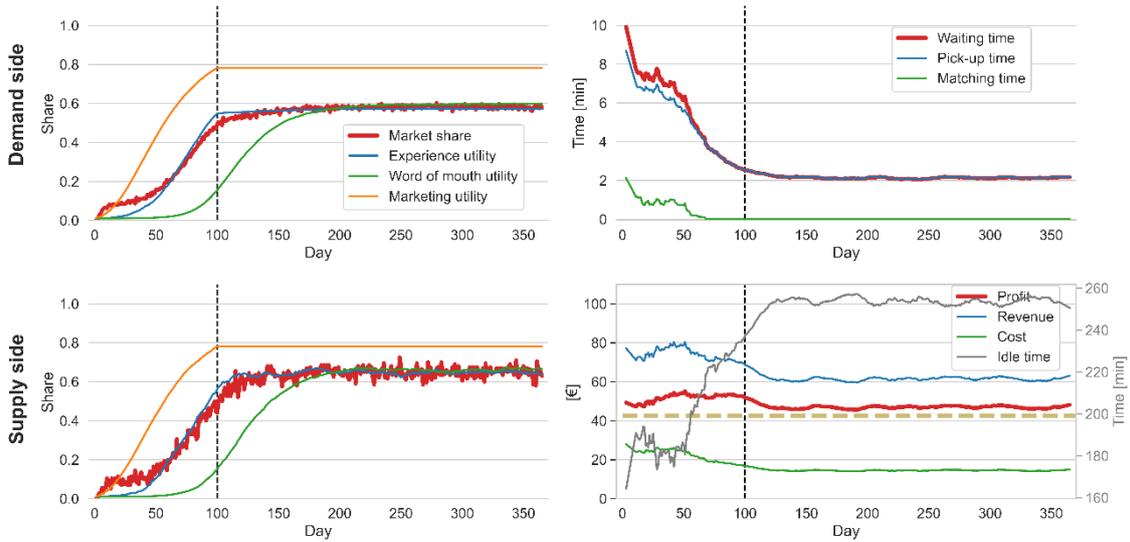

**Figure 2:** Single platform evolution with the baseline strategy. Platform applies marketing campaign and 40% discount on trip fares only in the initial 100 days (vertical dashed line), while commission rate is fixed to 10%.

*Correlation between platform alternative and the market shares*

Next, we introduce a second platform to the system and apply nested logit for the agents' mode choice. We investigate how the assumption of correlations in the nested choice model affects the equilibrium market shares. (fig. 3).

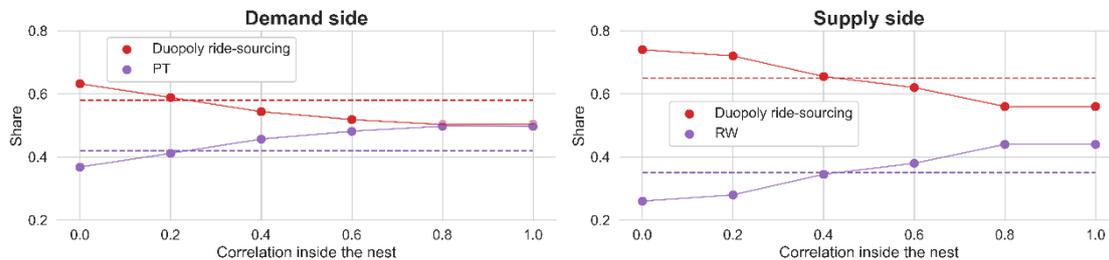

**Figure 3:** Ride-sourcing and PT market shares with different correlations between ride-sourcing alternatives. As correlation increases, total market share of ride-sourcing (both platforms) decreases. The dashed lines represent the ride-sourcing market share in the monopoly market with single platform.

*Three scenarios and relevant equilibria*

We fix the correlation rate inside the ride-sourcing choice nest at a moderate value of 0.4, and demonstrate the three scenarios with predetermined strategies as illustrated in fig. 3 to examine the emerging equilibria. In the first scenario, both platforms (P1, P2) apply baseline strategy, P2 launches 25 days later in the second scenario without any strategy change. In the third scenario, the P2 enter laters, but aggressively, with the 80% discount starting form day 25.

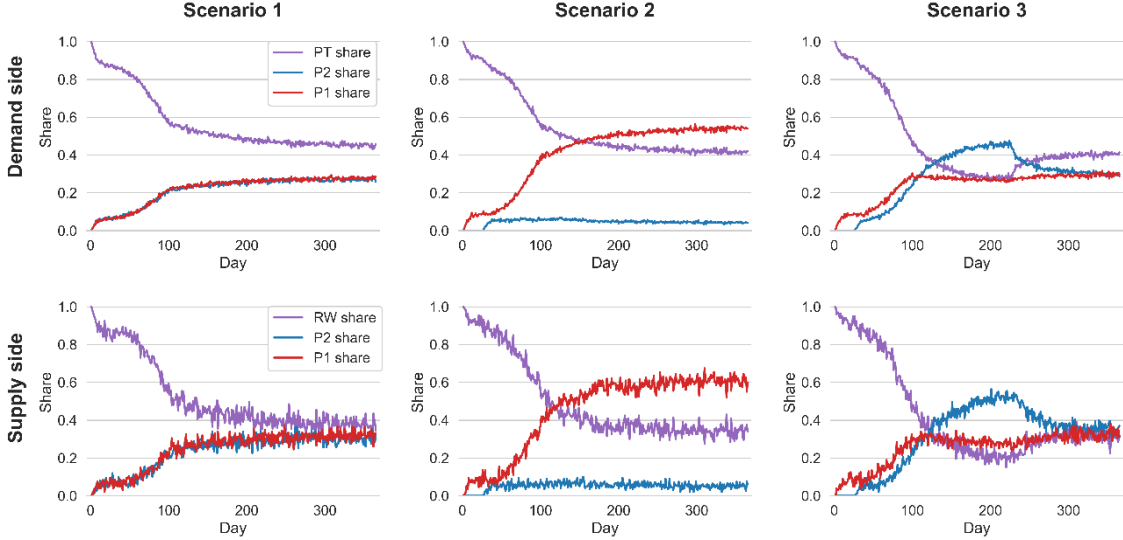

**Figure 3:** Three competition scenarios and resulting market shares.

Two platforms applying the same strategy, at the same time, end up with the same market shares (Scenario 1). However, late market entry with the same strategy results in failure (Scenario 2). This happens because agents have already started to use the early platform once P2 enters the market. Thus, P1 starts the cross-side network effects earlier, and as the utility and market share differences increase between two platforms, it becomes impossible for late platform to succeed. P2 requires an alternative strategy to compensate the late market penetration which means more subsidies on demand or/and supply side. In the Scenario 3, the late platform enters at the day 25 offering 80% discount for the next 200 days (instead of 100 days of 40% discount) on the demand side to overtake the early platform. In contrast to the second scenario, P2 reaches higher market share than P1 in the last scenario, at first (until day 225). Yet, as P2 terminates the discount its market share decreases and stabilizes (where it is supportable by network effects). Indeed, the market share bubble induced by disloyal agents, relying on discount, bursts with discount termination. While Scenarios 1 and 3 depict the market sharing regime, the second scenario resembles the winner-takes-all in the ride-sourcing market.

## 4. CONCLUSIONS

In this research, we shed light on the dynamics of ride-sourcing market in which two platforms compete with each other and the public transportation (for travellers) and reservation wage (for drivers). We use our day-to-day coevolutionary model featured by the S-shaped learning curves to capture the rise and fall of the system in MaaSSim. Our results underpin how correlation inside the ride-sourcing choice nest significantly affects the total market share of ride-sourcing platforms, which calls for further empirical studies. Assuming a moderate correlation rate, we analysed the competition in three different scenarios. We found that platforms with late entry to the market require more subsidies to trigger the cross-side network effects. However, subsidies can induce market share bubble for the platforms which can easily burst with the termination of subsidies, i.e., the platform stabilizes later, on the market share supportable by the network effects. We conclude that ride-sourcing market reaches an equilibria in long term, and both the "winner-takes-all" and the "market sharing" are the possible competition outcomes. Nevertheless, the market remains sensitive and late-entry alternatives may still reach significant market shares, mastering the network effects.

**ACKNOWLEDGEMENTS**

This research is funded by National Science Centre in Poland program OPUS 19 (Grant Number 2020/37/B/HS4/01847) and by Jagiellonian University under the program Excellence Initiative: Research University (IDUB).

**REFERENCES**

Ahmadinejad A., Nazerzadeh H., Saberi A., Skochdopole N., Sweeney, K. 2019. Competition in Ride-Hailing Markets. *SSRN Scholarly Paper*, No. 3461119.

Belleflamme P., Peitz M. 2018. Chapter 11: Platforms and network effects. *Game Theory and 8 Industrial Organization*, Cheltenham.

de Ruijter A., Cats O., Kucharski R., van Lint H. 2022. Evolution of labour supply in ridesourcing. *Transportmetrica B: Transport Dynamics*, Vol. 10, No. 1, p. 599–626.

Ghasemi F., Kucharski R. 2022. Modelling the Rise and Fall of Two-Sided Mobility Markets with Microsimulation *(arXiv:2208.02496). arXiv. https://doi.org/10.48550/arXiv.2208.02496*

Kucharski R., Cats O. 2022. Simulating Two-Sided Mobility Platforms with MaaSSim. *PLOS ONE*, Vol. 17, No. 6, p. e0269682.

Murre J. 2014. S-Shaped Learning Curves. *Psychonomic Bulletin & Review*, Vol. 21, No. 2, p. 344–356.

Rochet J.C., Tirole J. 2006. Two-Sided Markets: A Progress Report. The RAND *Journal of Economics*, Vol. 37, No. 3, 2006, pp. 645–667.

Shoman M., Moreno A. T. 2021. Exploring Preferences for Transportation Modes in the City of Munich after the Recent Incorporation of Ride-Hailing Companies. *Transportation Research Record*, Vol. 2675, No. 5, p. 329–338.

Siddiq A., Taylor T. A. 2022. Ride-Hailing Platforms: Competition and Autonomous Vehicles. *Manufacturing & Service Operations Management*, Vol, 24. No. 3, p. 1511–1528.

Zhang K., Nie Y. 2021. Inter-platform competition in a regulated ride-hail market with pooling. *Transportation Research Part E: Logistics and Transportation Review*, Vol. 151, p. 102327.